\documentclass[
aps,
prl,
reprint,
superscriptaddress,
nofootinbib,
nobibnotes
]{revtex4-2}

\usepackage{amsmath,amssymb,bm}
\usepackage{graphicx}
\usepackage{siunitx}
\usepackage{hyperref}
\usepackage{comment}

\begin{document}

\title{Higgsino Dark Matter Interpretation of the LUX-ZEPLIN\\ 248 keV Nuclear-Recoil Event}

\author{Katherine Freese}
\email{ktfreese@utexas.edu}
\affiliation{Weinberg Institute for Theoretical Physics, Department of Physics, The University of Texas at Austin, Austin, TX 78712, USA}
\affiliation{The Oskar Klein Centre, Department of Physics, Stockholm University, AlbaNova, SE-106 91 Stockholm, Sweden}
\affiliation{Nordita, Stockholm University and KTH Royal Institute of Technology, Hannes Alfvéns väg 12, SE-106 91 Stockholm, Sweden}
\author{Dionysios P. Theodosopoulos}
\email{d.theodosopoulos@utexas.edu}
\affiliation{Weinberg Institute for Theoretical Physics, Department of Physics, The University of Texas at Austin, Austin, TX 78712, USA}

\date{\today}

\begin{abstract}
We propose Higgsino dark matter as a potential interpretation of the $248~\mathrm{keV}$ nuclear-recoil event of interest reported by the LUX-ZEPLIN (LZ) experiment. LZ studied several rare background processes and detector effects in detail, but did not identify any as a likely explanation of the event. A nearly pure Higgsino naturally realizes inelastic dark matter through the off-diagonal $Z$ coupling of two nearly degenerate neutral Majorana states separated by a mass splitting $\delta$.
The same electroweak interaction fixes the inelastic Higgsino--nucleon scattering cross section, rather than leaving it as a free parameter.
We show that the predicted Higgsino inelastic scattering cross section approaches the published LZ two-sided 90\% confidence interval for a Higgsino mass $m_{\widetilde H}\sim1~\mathrm{TeV}$ and a mass splitting $\delta\sim350~\mathrm{keV}$.

\end{abstract}

\maketitle

\textit{Introduction---}
The nature of dark matter (DM) remains one of the longest oustanding unsolved problems in all of physics. Although observations indicate that it constitutes roughly 85\% of the mass in the Universe, its nature is as yet unknown. Weakly Interacting Massive Particles (WIMPs) are among the best  motivated candidates and arise in a variety of Beyond the Standard Model physics scenarios.

Direct detection laboratory experiments, originally proposed in the 1980s \cite{Goodman:1984dc,Drukier:1986tm},
seek to detect the scattering of WIMPs in our Galaxy as they
pass through terrestrial detectors situated in deep underground sites.
For decades the sensitivity of experiments has improved tremendously. The DAMA/LIBRA experiment observed an annual modulation consistent with a WIMP interpretation \cite{Bernabei:2013xsa}, but recent experiments made of the same NaI material are in tension with the result \cite{ANAIS-112:2025fne}.
Recently, the LUX-ZEPLIN (LZ) Collaboration 
performed a search for WIMP dark matter particles interacting with xenon nuclei with a $2.84$ tonne-year exposure over an extended nuclear-recoil energy range, $5.4$--$270~\mathrm{keV}$ \cite{LZ:2026}.  
The higher recoil range with large exposure allowed for searches for 
WIMP-nucleon interactions in a regime not previously studied.  
In a region with very low expected backgrounds, they found an event of interest, consistent with a nuclear recoil of 248 $\pm$ 23 (stat) $\pm$ 23 (sys) keV.  Using a profile likelihood ratio test, they found the background only hypothesis to be in tension at a global significance of 2.6$\sigma$ with a maximum local significance of 3.4$\sigma$ across the models they tested.
LZ reports that several rare background processes and detector effects were studied in detail, but none was identified as a likely explanation for the event of interest. While this significance is far below the threshold required for a discovery claim, the absence of an identified likely background motivates considering well-motivated particle-physics interpretations of the event.

Supersymmetric (SUSY) extensions of the Standard Model (SM) of particle physics readily predict new particles that could serve as WIMPs.
In this Letter, we propose Higgsino dark matter as a potential explanation for the LZ event.
The Higgsino is the SUSY partner of the Higgs boson of the SM.
  
The high nuclear recoil energy accessed by LZ, up to $270\ \mathrm{keV}$, 
opens up their study to new interactions, beyond the standard Spin Independent (SI) elastic scattering whose nuclear recoil spectrum falls rapidly above $\sim50~\mathrm{keV}$
even for TeV WIMPs.  
Although LZ did previously publish results which went up to these high nuclear recoil energies \cite{LZ:2023lvz}, 
their new results have significantly higher exposure, due to 220 live days of data used in the current analysis vs. 60 days in their earlier paper.
In the new data, they studied two cases: 1) elastic WIMP-nucleon scattering described by 20 fully relativistic covariant Lagrangians, in their nonrelativistic limits; and 2) inelastic scattering interactions in which the WIMP transitions into a more massive state.
Both of these possibilities are interesting in the context of high energy recoil spectra (with lower energy components suppressed).
In particular, LZ searched for high-energy nuclear recoils arising due to elastic scattering with all 20 possible interaction types, as well as inelastic interactions arising from spin-independent $(\mathcal{O}_1)$ and spin-dependent $(\mathcal{O}_4)$ operators.\footnote{Here, $\mathcal{O}_1$ and $\mathcal{O}_4$ refer to the Non-Relativistic Effect Field Theory (NREFT) operators that can be used to describe WIMP/nucleon interactions.}

We consider Higgsino dark matter as a well-motivated realization of the inelastic scattering scenario considered by LZ \cite{LZ:2026, Graham:2024syw}.
The Higgsino interpretation is particularly predictive because the relevant interaction strength is not a free parameter. For a nearly pure Higgsino, the off-diagonal $Z$ coupling fixes the inelastic Higgsino--nucleon cross section. 
The mass splitting $\delta$ therefore largely determines whether the interaction is kinematically accessible in xenon. 
The 1 TeV inelastic case studied by LZ corresponds to the Higgsino mass which approximately gives the correct relic abundance today to explain the dark matter via thermal production in the early Universe.
In this paper, we specifically compare the prediction for the case of the thermal $1.1~\mathrm{TeV}$ Higgsino with the LZ two-sided confidence interval and show that the event of interest can be accommodated by inelastic Higgsino dark matter particles.
In a followup paper, we will generalize to other (heavier, nonthermal) Higgsino masses and again show agreement with LZ data.

\textit{Higgsino dark matter---}
A particularly well-motivated realization of inelastic dark matter is the 
Higgsino.
one of the remaining classic supersymmetric (SUSY) WIMP candidates
\cite{Krall:2017xij,Co:2021ion,Bhattiprolu:2025zwt,Kowalska:2018toh}.
In the minimal supersymmetric Standard Model (MSSM), the up- and down-type Higgsino doublets form a Dirac fermion with mass $\mu$ before electroweak symmetry breaking (EWSB). 
After EWSB, interactions with the Standard Model Higgs split the neutral Dirac Higgsino into two Majorana fermions \cite{Nagata:2014wma},
$\widetilde H_1$ and $\widetilde H_2$, with masses $m_{\widetilde H_{1,2}}\approx\mu \mp \frac{\delta}{2}$, where $\delta$ is the Majorana mass splitting given by
\begin{equation}
    \delta
    \simeq
    m_Z^2
    \left(
        \frac{\sin^2\theta_W}{M_1}
        +
        \frac{\cos^2\theta_W}{M_2}
    \right).
    \label{eq:higgsino_splitting}
\end{equation}
Here $m_Z$, $M_1$, and $M_2$ denote the Z boson, bino and wino masses, respectively, and $\theta_W$ is the Weinberg angle.
In the heavy-gaugino limit ($M_{1,2}\gg m_Z,\mu$), the two neutral Higgsinos constitute a narrowly split pseudo-Dirac pair, $\delta\ll\mu$.
The scenario that the gauginos and other superpartners lie well above the weak scale is consistent with the absence of new physics at the LHC 
\cite{Arkani-Hamed:2004ymt,Giudice:2004tc,Gonski:2025wzh}.

For a mass splitting of a few hundred keV, Eq.~\eqref{eq:higgsino_splitting} implies a characteristic electroweak gaugino mass scale of order $10^7\ \mathrm{GeV}$. Such heavy gauginos exceed naturalness expectations and entail electroweak fine-tuning in the MSSM, leaving a residual ``little hierarchy problem" \cite{Baer:2015rja}, 
a theoretical cost of the required SUSY spectrum.

If the lighter Higgsino $\widetilde H_1$ is the lightest supersymmetric
particle, it can be stable and serve as a WIMP dark-matter candidate \cite{Ellis:1983ew}. 
The heavier state $\widetilde H_2$ decays on a timescale much shorter than the age of the Universe, so present-day dark matter consists of $\widetilde H_1$.
For a mass $m_{\widetilde H}\simeq 1.1~{\rm TeV}$, it obtains the observed dark-matter abundance $\Omega h^2=0.12$ through standard thermal freeze-out \cite{Kowalska:2018toh}. 
For heavier masses, Higgsinos would be overproduced and their abundance must be diluted with nonstandard thermal histories.

The neutral-current interaction is off diagonal in the Higgsino mass
basis, and hence tree-level $Z$ exchange induces inelastic scattering---the mass of the initial and the final particle state differs by $m_{\widetilde{H}_2}-m_{\widetilde{H}_1}=\delta$
\cite{Tucker-Smith:2001myb}.
The interaction is predominantly coherent and neutron coupled \cite{Nagata:2014wma,Kowalska:2018toh,Essig:2007az}. 
The corresponding Higgsino--nucleon cross section (for vector coupling) is 
\begin{equation}
    \sigma_{\widetilde H N}
    =
    \frac{G_F^2\mu_N^2}{8\pi}
    \simeq
    1.86\times10^{-39}~{\rm cm}^2 ,
    \label{eq:higgsino_neutron_xsec}
\end{equation}
where $\mu_N$ is the Higgsino--nucleon reduced mass.
This interaction strength is many orders of magnitude larger than the elastic cross sections currently probed by conventional direct-detection experiments. The Higgsino nevertheless remains viable because the scattering is inelastic; the lighter $\widetilde H_1$ needs to convert to $\widetilde H_2$. As $\delta$ increases, only particles in the high-velocity tail  of the WIMP distribution in the halo of our Galaxy have sufficient kinetic energy to excite the heavier state.

Although elastic Higgsino--nucleon scattering can arise through tree-level Higgs exchange or electroweak loop processes, it is suppressed. In the heavy-gaugino limit, the Higgsino--Higgs coupling is induced only through gaugino mixing and is therefore suppressed by $m_Z/M_{1,2}$ \cite{Krall:2017xij}. The remaining one-loop elastic contribution is also strongly suppressed by cancellations among diagrams for the observed Higgs mass \cite{Hisano:2011cs,Hisano:2012wm}, leaving the elastic cross section below the neutrino floor.

\textit{Comparison with the LZ result---} LZ reports two-sided 90\% confidence-level intervals on the SI scalar inelastic
DM--nucleon scattering cross section $\sigma_{\rm SI}^N$ as a function of the mass splitting
$\delta$ for $m_\chi=1~\mathrm{TeV}$ \cite{LZ:2026}. The published limits assume scalar
coupling normalization. For SI vector coupling, which is relevant in our case, LZ provides the conversion
\begin{equation}
    \sigma_V^N
    =
    \left[
    \frac{A}
    {(A-Z)-(1-4\sin^2\theta_W)Z}
    \right]^2
    \sigma_{\rm SI}^N
    \simeq
    3.2\,\sigma_{\rm SI}^N ,
    \label{eq:vector_conversion}
\end{equation}
for xenon. Since the Higgsino interacts through $Z$-mediated vector
couplings, we multiply the published LZ cross-section interval by this
factor and compare it directly with the Higgsino--nucleon cross section,
$\sigma_{\widetilde H N}$.

\begin{figure}[t]
    \centering
    \includegraphics[width=\columnwidth]
    {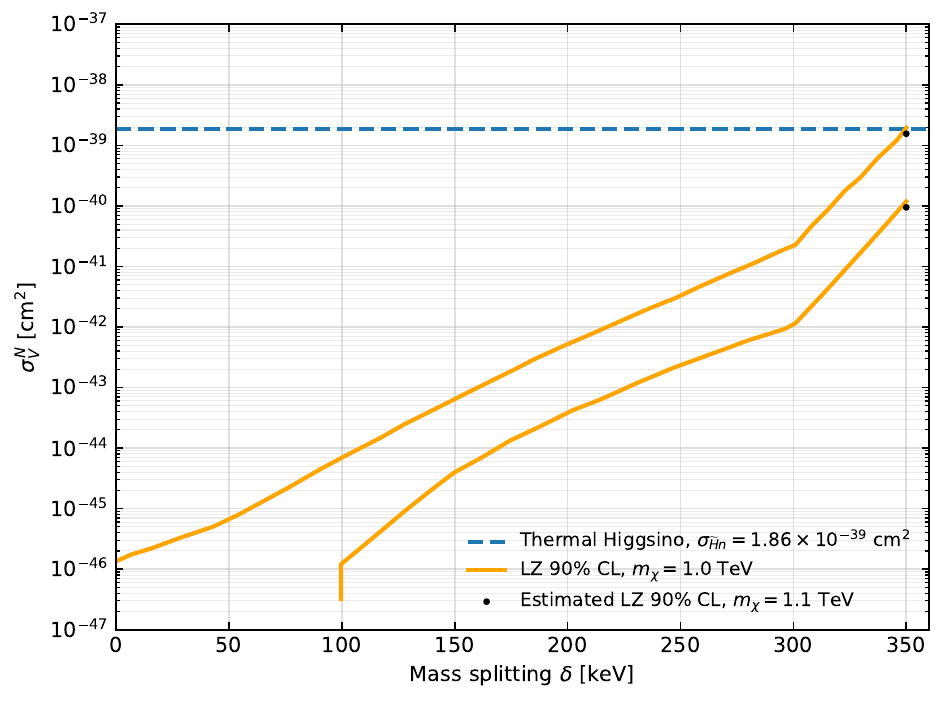}
    \caption{
    Two-sided 90\% confidence-level interval on the SI inelastic DM--nucleon cross section reported by LZ for $m_\chi=1~\mathrm{TeV}$, expressed in the vector-coupling normalization using the LZ conversion $\sigma_V^N\simeq3.2\,\sigma_{\rm SI}^N$ for xenon. The horizontal line shows the tree-level Higgsino--nucleon cross section, $\sigma_{\widetilde{H} N}\simeq1.86\times10^{-39}~\mathrm{cm}^2$. Near the upper end of the published scan, the Higgsino prediction lies within the LZ two-sided 90\% confidence interval.
    The two black points at $\delta=350~\mathrm{keV}$ show our estimated lower and upper limits of the two-sided 90\% confidence interval for $m_\chi=1.1~\mathrm{TeV}$, obtained by rescaling the published $m_\chi=1~\mathrm{TeV}$ LZ limits using the efficiency-weighted recoil rate; the black points are lower than the orange curves by a factor of 0.8.
    }
    \label{fig:lz_higgsino}
\end{figure}

Figure~\ref{fig:lz_higgsino} shows the resulting comparison. 
The Higgsino cross section enters the two-sided LZ 90\% confidence interval at large mass splitting. 
In particular, at
$\delta=350~\mathrm{keV}$ the vector-normalized LZ interval is
approximately
\begin{equation}
    1.09\times10^{-40}~\mathrm{cm}^2
    \lesssim
    \sigma_V^N
    \lesssim
    1.92\times10^{-39}~\mathrm{cm}^2 ,
    \label{eq:lz_interval_1tev}
\end{equation}
which contains the Higgsino prediction
$\sigma_{\widetilde H N}$.
Thus, for $m_{\widetilde H}=1~\mathrm{TeV}$ and $\delta=350~\mathrm{keV}$, the Higgsino prediction lies within the published LZ two-sided 90\% confidence interval.  

The comparison above is made at the benchmark mass, $m_\chi=1~\mathrm{TeV}$, for which LZ reports the published two-sided interval. For a thermal Higgsino, however, the relic abundance selects a slightly larger mass, $m_{\widetilde H}\simeq1.1~\mathrm{TeV}$. We therefore next estimate the impact of this modest mass shift on the LZ limits.

Increasing $m_\chi$ reduces the local number density, $n_\chi=\rho_\chi/m_\chi$, but also increases the DM--nucleus reduced mass, and therefore lowers the minimum velocity required for inelastic scattering. The latter effect can be important at $\delta=350~\mathrm{keV}$, where the rate is dominated by the high-velocity tail of the DM distribution. 
Increasing $m_\chi$ changes both the normalization and shape of the recoil spectrum. A rigorous treatment of the latter would require repeating the LZ likelihood analysis for
$m_\chi=1.1~\mathrm{TeV}$. 
We therefore estimate the mass dependence by retaining only the change in the total expected recoil rate, while neglecting the associated change in spectral shape.

To quantify this effect, we rescale the two-sided limits in Fig.~\ref{fig:lz_higgsino} at $\delta=350~\mathrm{keV}$ using the ratio of the efficiency-weighted nuclear-recoil rates $R(m_\chi,\delta)$, defined per unit detector mass and exposure time, for $m_\chi=1.0$ and $1.1~\mathrm{TeV}$. We calculate the recoil spectra with \textsc{WimPyDD} \cite{Jeong:2021bpl}, adopting the same astrophysical assumptions as in the LZ analysis
\cite{Baxter:2021pqo}, and define the
efficiency-weighted rate as
\begin{equation}
    R(m_\chi,\delta)
    =
    \int dE_R\,
    \epsilon_{\rm LZ}(E_R)
    \frac{dR}{dE_R}(m_\chi,\delta),
    \label{eq:recoil_rate}
\end{equation}
where $\epsilon_{\rm LZ}(E_R)$ is the total nuclear-recoil efficiency reported by LZ. We use the same interaction strength for both masses, so its normalization cancels in the rate ratio.

Neglecting changes in the recoil-spectrum shape, we rescale the cross-section limits at $\delta=350\ \mathrm{keV}$ with the ratio
\begin{equation}
    \mathcal R_{350}
    \equiv
    \frac{R(1.0~\mathrm{TeV},350~\mathrm{keV})}
         {R(1.1~\mathrm{TeV},350~\mathrm{keV})}
    \approx0.8,
    \qquad
    \sigma_{\rm lim}^{1.1}
    \simeq
    \mathcal R_{350}\,
    \sigma_{\rm lim}^{1.0}.
    \label{eq:mass_rescaling}
\end{equation}
The two points at $\delta=350~\mathrm{keV}$ in
Fig.~\ref{fig:lz_higgsino} show the lower and upper limits obtained from
this estimate. The resulting interval lies below the Higgsino prediction,
so the thermal Higgsino falls outside our estimated
$m_\chi=1.1~\mathrm{TeV}$ 90\% confidence interval at
$\delta=350~\mathrm{keV}$.

This rescaling should, however, be regarded only as an estimate.
Changing the DM mass modifies not only the total rate but also the recoil spectrum, whereas the LZ confidence interval is obtained from a profile-likelihood analysis that uses the full signal distribution.
A dedicated LZ likelihood analysis is therefore required to derive rigorous constraints at $m_\chi=1.1~\mathrm{TeV}$.

Importantly, $\delta=350~\mathrm{keV}$ is near the maximum kinematically accessible splitting by xenon detectors \cite{Tucker-Smith:2001myb,Barello:2014uda}. 
Therefore, for $\delta$ slightly above $350~\mathrm{keV}$, the inelastic scattering rate is suppressed and thus the two-sided curves in Fig.~\ref{fig:lz_higgsino}
are expected to increase significantly to higher values of the cross-section. 
A slight increase in the mass splitting will therefore
bring the thermal 1.1 TeV Higgsino into the two-sided LZ 90\% confidence interval.\footnote{Note that if one takes very seriously the discrepancy between Higgsino mass of 1.1 vs. 1.0 TeV in the LZ data as well as the computation of dark matter relic density, the 1.0 TeV mass higgsino would make up $\sim 80\%$ of the dark matter.}


\textit{Discussion---}
In this Letter, we have explored a Higgsino dark-matter interpretation of the $248\pm23\,(\mathrm{stat})\pm23\,(\mathrm{syst})~\mathrm{keV}$ nuclear-recoil event of interest observed by LZ \cite{LZ:2026}. The event lies in a region of low expected background, and none of the rare background processes or detector effects investigated by the LZ Collaboration was identified as a likely explanation.
The event of interest can be interpreted by Higgsino dark matter with a mass $\sim\mathrm{TeV}$, and a mass splitting $\sim350~\mathrm{keV}$.

The proximity of the Higgsino prediction to the LZ preferred region is particularly interesting because the interaction strength is not chosen to reproduce the event.
For a nearly pure Higgsino, the off-diagonal $Z$ interaction fixes the inelastic Higgsino--nucleon cross section to $\sigma_{\widetilde H N}\simeq1.86\times10^{-39}~\mathrm{cm}^2$, while standard thermal freeze-out independently selects a Higgsino mass $m_{\widetilde H}\simeq1.1~\mathrm{TeV}$. The mass splitting $\delta$ therefore provides the principal parameter controlling whether this otherwise large electroweak interaction is kinematically accessible in a xenon detector.
It is thus nontrivial that, for splittings near $\delta\sim350~\mathrm{keV}$, the fixed Higgsino prediction approaches the two-sided confidence interval associated with the LZ event of interest. 

In a follow-up paper to this Letter, we will present a likelihood analysis extending to 
heavier  Higgsino masses and larger mass splittings; we will show that heavier (nonthermal) Higgsinos can be consistent with LZ and other data.

\textit{Acknowledgments---}
KF is grateful for support from the Jeff \& Gail Kodosky Endowed Chair in Physics at the University of Texas. KF acknowledges support from the Swedish Research Council (Contract No. 638-2013-8993).  KF and DT acknowledge support by the U.S. Department of Energy, Office of Science, Office of High Energy Physics program under Award Number DE-SC-0022021.

\bibliography{refs}

\end{document}